\documentclass{aa}
\usepackage{natbib}
\bibpunct{(}{)}{;}{a}{}{,}
\usepackage{psfig}
\usepackage{epsfig}

\newcommand\sax{BeppoSAX}

\newcommand\src{G327.1--1.1}

\newcommand\E[1]{\times10^{#1}}
\newcommand\dash{\hbox{\rm --}}
\newcommand\gsim{\;\lower4pt\hbox{${\buildrel\displaystyle >\over\sim}$}\,}
\newcommand\lsim{\;\lower4pt\hbox{${\buildrel\displaystyle <\over\sim}$}\,}
\newcommand\approxgt{\mathrel{\hbox{\rlap{\lower.55ex \hbox {$\sim$}}
        \kern-.3em \raise.4ex \hbox{$>$}}}}
\newcommand\approxlt{\mathrel{\hbox{\rlap{\lower.55ex \hbox {$\sim$}}
        \kern-.3em \raise.4ex \hbox{$<$}}}}

\newcommand\hh{\hbox{$^{\rm h}$}}
\newcommand\mm{\hbox{$^{\rm m}$}}
\newcommand\sss{\hbox{$^{\rm s}$}}

\newcommand\U[1]{{\,\rm #1}}
\newcommand\kms{km\,s^{-1}}
\newcommand\cmq{cm^{-2}}
\newcommand\cmc{cm^{-3}}
\newcommand\ergs{erg\,s^{-1}}

\newcommand\rs[1]{_\mathrm{#1}}
\newcommand\nh{N\rs{H}}
\newcommand\mH{m\rs{H}}
\newcommand\Msun{M_\odot}
\newcommand\Tsh{T\rs{sh}}
\newcommand\TkeV{T\rs{keV}}
\newcommand\Vsh{V\rs{sh}}
\newcommand\tSNR{t\rs{SNR}}
\newcommand\nz{n\rs{0}}
\newcommand\ESN{E\rs{SN}}
\newcommand\Mtot{M\rs{tot}}
\newcommand\dE{d\rs{E}}
\newcommand\VshS{V\rs{sh,S}}
\newcommand\tSNRS{t\rs{SNR,S}}
\newcommand\nzS{n\rs{0,S}}
\newcommand\ESNS{E\rs{SN,S}}
\newcommand\MtotS{M\rs{tot,S}}
\newcommand\dES{d\rs{E,S}}
\newcommand\RshS{R\rs{sh,S}}
\newcommand\EMS{\hbox{EM}\rs{S}}
\newcommand\TeffS{T\rs{eff,S}}
\newcommand\Rsh{R\rs{sh}}
\newcommand\EM{\hbox{EM}}
\newcommand\Teff{T\rs{eff}}
\newcommand\Mx{M\rs{x}}
\newcommand\bRsh{\widetilde{R}\rs{sh}}
\newcommand\bEM{\widetilde{\hbox{EM}}}
\newcommand\bTeff{\widetilde{T}\rs{eff}}
\newcommand\bMx{\widetilde{M}\rs{x}}
\newcommand\cooleff{\epsilon\rs{c}}
\newcommand\nsav{\langle n^2\rangle}
\newcommand\Lrad{L\rs{1GHz}}

\newcommand\gm{\gamma}
\newcommand\sg{\sigma}

\begin{document}


\title{\sax\ observation of the composite remnant \src}

      \author{
        F. Bocchino\inst{1}
         \and
        R. Bandiera\inst{2}
}
\offprints{\\ F. Bocchino (bocchino@astropa.unipa.it)}

\institute{
       Osservatorio Astronomico di Palermo, Piazza del Parlamento 1,
       90134 Palermo, Italy
\and
       Osservatorio Astrofisico di Arcetri, Largo E. Fermi 5,
       I-50125 Firenze, Italy
}

\date{Received;Accepted}

\abstract{We report an X-ray study of the composite supernova remnant
\src, with particular emphasis on its thermal emission. By virtue of a
combined spatial and spectral analysis, we have been able to model the
X-ray emission of the remnant as a sum of two components: a non-thermal
component, due to the pulsar nebula and the pulsar itself, and a
thermal component, of which we have analysed spectrum and morphology,
after proper subtraction of the plerion. We discuss three possible
interpretations of the thermal emission of \src: pure Sedov expansion,
expansion through a
inhomogeneous medium with evaporation from ISM clouds, and radiative
expansion. On the light of our new data and
interpretation, we have re-derived all the physical parameters of this SNR. In
the framework of Sedov or radiative expansion we derive a longer age
than previously estimated ($1.1\times 10^4$), thus not requiring a high
velocity for the pulsar.
}

\maketitle

\markboth{F. Bocchino and R. Bandiera}{\sax\ observation of \src.}

\section{Introduction}

The class of supernova remnants (SNRs) contains a wide variety of objects, that
cannot be accounted for by the traditional subdivision in just two categories,
namely shell-like remnants (powered by a blast wave expanding in the ambient
medium) and filled-center (or Crab-like, or plerionic) remnants (powered by a
spinning-down pulsar).
The class of composite SNRs has in fact been introduced (\citealt{hb87}) in
order
to arrange a number of objects that neither could be classified as pure
shell-like nor as pure filled-center remnants.
This classification was originally only morphological, by requiring just the
coexistence of a shell and of a center-filled component: the underlying belief
was that
the inner component always has a plerionic nature.
Such scenario, originally devised for objects in which both components are
detected in radio,
seemed to apply also to those appearing as pure shells in radio, but with
a centrally peaked emission in X rays.
A natural explanation was that in these SNRs the central X-ray component is due
to synchrotron emission of plerionic nature, which happens to be too weak in
radio.

From the analysis of the X-ray spectra however it became clear that for
a large fraction of composite SNRs (the so-called ``mixed-morphology''
remnants) the emission from the central component is thermal.  Along
this line (\citealt{wl91}, hereafter WL) presented a model showing that
a centrally peaked (thermal) X-ray emission characterizes SNRs in which
the blast wave is expanding in a cloudy interstellar medium and induces
a delayed evaporation of the clouds.  This model has been applied to
several SNRs, although its validity is still under debate.  For
instance \citet{rps94} found that the radial profiles of SNR W44 are
nicely fitted by a WL model; but \citet{hhs97} then showed that this
model fails in dating the SNR, and suggested an alternative scenario
based on the idea that the shock has reached the radiative phase.  A
more detailed radiative model for W44 has been recently presented
(\citealt{csm99,scm99}), which includes thermal conduction and a
density gradient in the ambient medium.  An alternative suggestion
(\citealt{pet01}) is that even a shell-type SNR which expands in a
strong density gradient may mimic a center-filled morphology, if seen
with a peculiar orientation.  Further investigation is still required
to clarify which scenario is most appropriate to describe the majority
of thermal X-ray SNRs with a center-filled morphology.

On the observational side, the diagnostics and even the recognition of
mixed-morphology SNRs are not easy tasks.
The determination of the nature (thermal or non-thermal) of the central X-ray
component on the basis of its X-ray spectrum is in principle straightforward.
However a thermal and a non-thermal component may coexist in the same remnant,
in which case both a high spatial resolution and a broad-spectrum sensitivity
are required in order to study them separately.
The coexistence, in composite SNRs, of more components typically implies
a dynamical interaction between them, a study of which may provide
important clues on the physical conditions.
For instance,
an X-ray spectral determination of the pressure in the shell component,
together with the requirement of a pressure balance between the thermal remnant
and the plerion, allows an estimate of the plerionic magnetic field.

Although there are various composite SNRs that in X rays show both a shell-like
and a plerionic component, up to now only in one case (W44) a plerionic
component is clearly seen coexisting with a centrally peaked thermal X-ray
component.
In this paper we suggest that also SNR \src\ may present
such coexistence, and be in
various respects similar to W44.

\src\ is a composite SNR with a rather complex structure.  Details of
its morphology have been revealed only recently at radio wavelengths,
by the 843~MHz MOST SNR survey (\citealt{wg96}).  In particular its
plerionic component is substantially offset with respect to the
centroid of an otherwise rather symmetric remnant.
 A recent Chandra observation has in fact confirmed the presence of a
compact source (P.O.~Slane, private communication) on the tip of a
finger-like feature extending from the radio plerion to the north-west
direction, which also coincides with a hard X-ray source
(\citealt{swc99}, hereafter SWC).  On the other hand, the thermal
component of \src\ has not been mapped in detail, and its origin is still
uncertain. SWC interpreted it as a
thermal shell, and applied the Sedov solution to derive the shell
parameters. However, the presence of the strong non-thermal nebula
prevented them from showing the putative limb-brightened X-ray morphology
of the thermal emission.  In this work, we attempt to derive the
spatial morphology of the thermal component.

The plan of the paper is the following: in Sect.~2 the observation is
described; Sect.~3 presents results from a joined spectral and spatial analysis
of the data; Sect.~4 is devoted to a discussion on the nature of \src; Sect.~5
concludes.

\section{Observations}

The \sax\ imaging instruments are the Medium-Energy Concentrator Spectrometer
(MECS; 1.8\dash10~keV; \citealt{bbp97}) and the Low-Energy Concentrator
Spectrometer (LECS; 0.1\dash10~keV; \citealt{pmb97}).
The MECS consists of two grazing incidence telescopes with imaging gas
scintillation proportional counters in their focal planes.
The LECS uses an identical concentrator system as the MECS, but utilizes an
ultra-thin entrance window and a driftless configuration to extend the
low-energy response to 0.1~keV.
The fields of view (FOV) of the LECS and MECS are circular with diameters of
37\arcmin\ and 56\arcmin, respectively.
In the overlapping energy range, the space resolution of both instruments is
similar and corresponds to 90\% encircled energy within a radius of 2\farcm5 at
1.5~keV.
In addition, the \sax\ payload includes two high energy instruments - the High
Pressure Gas Scintillation Proportional Counter (HPGSPC; 5\dash120~keV;
\citealt{mgs97}) and the Phoswich Detection System (PDS; 15\dash300~keV;
\citealt{fcd97}).

\begin{figure}
\centerline{\epsfig{file=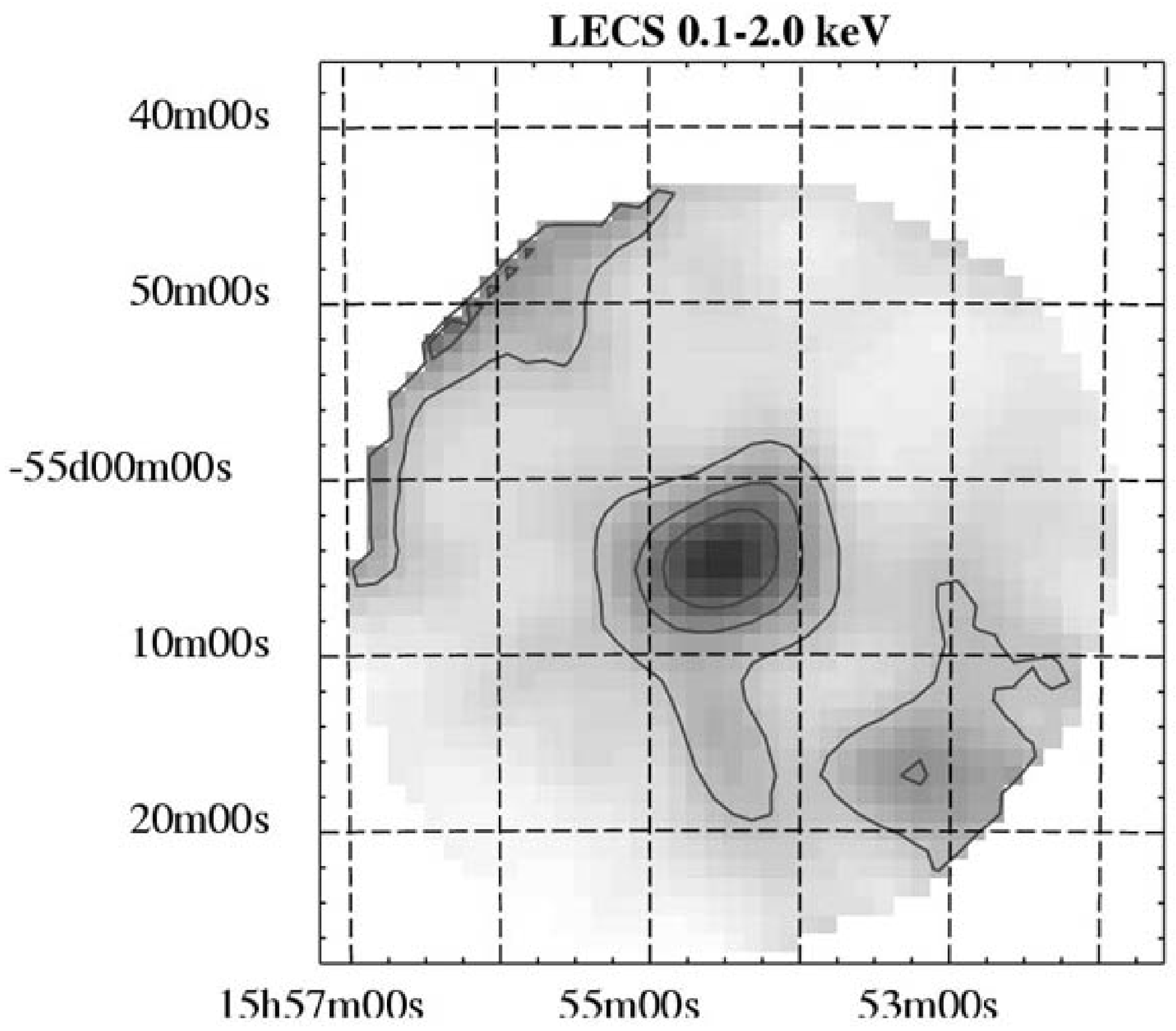,width=9.0cm}}
\centerline{\epsfig{file=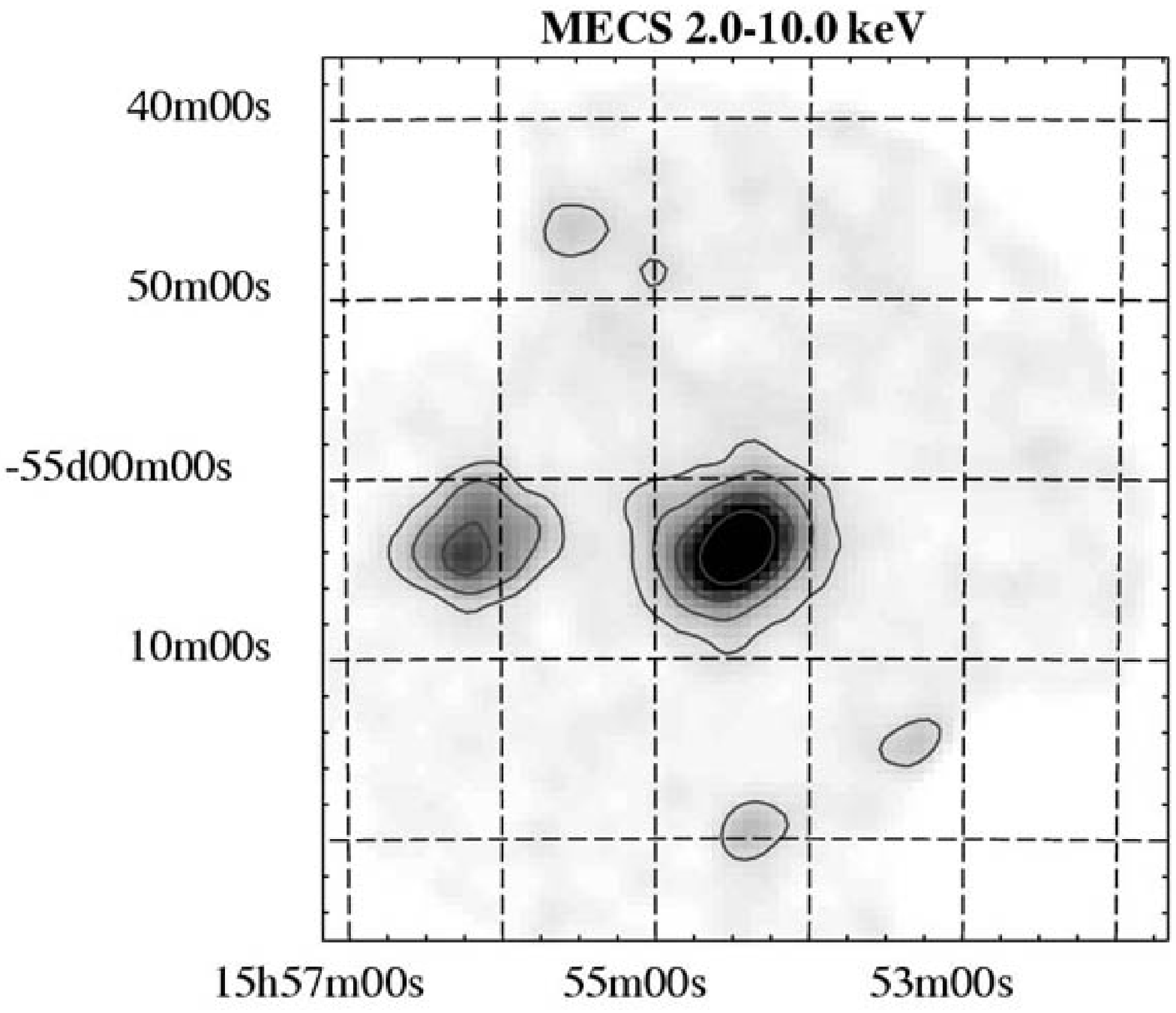,width=9.0cm}}
\caption{ 0.1\dash2 keV LECS image (top) and 2\dash10 keV MECS image
(bottom) of the region of \src.  Pixel size is 64\arcsec\ and
32\arcsec\ respectively, and a smoothing of 2 pixels has been
applied.  The images are exposure, vignetting and background
corrected.  LECS contours are at 0.7, 1.3 and $1.8\times
10^{-4}\U{cnt\,pix^{-1}}$, while MECS contours are at 1/16, 1/8, 1/4
and 1/2 of the peak value ($1.2\E{-3}\U{cnt\,pix^{-1}}$).}

  \label{mecs}
\end{figure}

The region of sky containing \src\ was observed between 1999 March 14 and March
16.
The pointing direction was R.A.=$15\hh54\mm30\sss$,
Dec.=$-55\degr04\arcmin36\arcsec$ (J2000).
In order to avoid solar scattered emission and other contaminating effects,
data were limited to intervals when the elevation angle above the
Earth's limb was $>4\degr$ and when the instrument configurations were
nominal.
The screened exposures in the LECS and MECS are 31.0~ks and 82.5~ks,
respectively.
The LECS and MECS images are shown in Fig.~\ref{mecs}.
\src\ is located, in both images, near the field center, as a resolved peak
centered near R.A.=$15\hh54\mm29\sss$, Dec.=$-55\degr03\arcmin48\arcsec$.
(J2000).

An unresolved peak not related to \src, located at position R.A.=$15\hh56\mm09\fs6$,
Dec.=$-55\degr03\arcmin11\arcsec$, with a hard power-law spectrum, is also
present in MECS2 data.
However it neither appears in LECS nor in MECS3 data, nor it is present in
existing catalogs (e.g.\ RASS).
Since the spectral response of MECS2 and MECS3 are similar, if this source were
real the only way to miss it in MECS3 data would be to have it right behind a
rib of the strongback: however only soft sources can be hidden completely in
this way; while ribs are transparent above 5~keV.
Of course it cannot be an effect of source variability, because the
integrations of the two MECS are virtually contemporaneous.
Moreover in a nearby archival \sax\ observation (Target: HD141926, Obs.Code:
20309004), this source again appears in MECS2, while it does not in MECS3.
The conclusion we have reached is that the peak seen in MECS2 is a ghost
image.
It cannot be due to problems in the electronics, because it is also visible in
the archival MECS2 image, while tests on dark and bright earth exposure close
in time to our observation do not reveal anything abnormal.
The peak is presumably produced by a bright source located off-axis (but that
has not been identified), and
therefore we have simply not considered it in the further analysis.
Anyway it does not seem to affect the data in the direction of \src.
We thank F.~Fiore and G.~Cusumano, of the \sax\ team, for their valuable advice
during the analysis of the nature of this source.

\section{Results}

\subsection{Spectral analysis}

The thermal and non-thermal components of \src\ are only partially resolved by
\sax, and therefore an accurate analysis is required to separate them.
This analysis should take advantage of the differences in the spectra of
the (soft) thermal and (hard) non-thermal components, as well as of the
differences in their position and size.

We have performed spectral fits using different spatial and spectral
selections of the available photons, and combined the results.
In all cases we exclude photons with energies which are outside the ranges
0.1\dash8~keV for LECS and 1.6\dash10.5~keV for MECS.
Fittings have used, separately or in combination, the following two emission
components: a non-thermal
component, modelled as a power law with photon index $\gm$,
and a thermal one, modelled as
due to ionized plasma hit by the shock wave ({\sc mekal}
model in XSPEC V11.0).
Reference abundances are from \citealt{ag89}, while
the interstellar absorption has been modelled using the \citet{mm83}
cross-sections.
The background was taken in a sector located to NE of \src, with radial
distances ranging from 10\arcmin\ to 20\arcmin (an area reasonably free from
point-like sources), and then corrected for vignetting and extraction areas.

We have first extracted a spectrum from a circular region with
$3\arcmin$ radius, centered on the X-ray peak position
(hereafter referred to as the
``plerion" region), with the aim of minimizing the contribution of the thermal
component.
For the spectral fitting we have generated an
``ad-hoc" response file which takes into account the source extension, that has
been estimated in the following way.
In the 2--10 keV MECS band the apparent size of
the source, as derived with a gaussian fit, is $3.6\arcmin$ (FWHM),
while that of the MECS Point spread Function (PSF, obtained by a
similar fit to the archive observation of Cyg X-1) in the same energy range
is $2.4\arcmin$: therefore the intrinsic size of the source can be estimated
as $\sqrt{3.6^2-2.4^2}=2.7$\arcmin\ (FWHM), with an uncertainty 
$\sim$0.3\arcmin. 

We have performed different fits (labelled as Fit A, B and C) to the spectrum
of the plerion region, whose results are reported in Table~\ref{fitres}.
A pure power-law model using the LECS+MECS data (Fit A) provides a good
overall fit of the spectra in terms of $\chi^2$ test, but
tends to underestimate the flux below 1 keV. The addition of a
thermal component is needed in order to have the spectra well fitted over the
whole LECS+MECS bandwidth (Fit B). It is worth noting that both the value of the
absorption and that of the power-law index are significantly different from the
values obtained with the pure power-law model. The results of Fit B are
confirmed by a pure power-law fit which uses only the 2--10 keV data of both 
instruments (Fit C), where the contribution of the thermal component is minor.
Therefore we have decided to take the parameters of Fit B as our best guess
for the plerion.

\begin{figure}
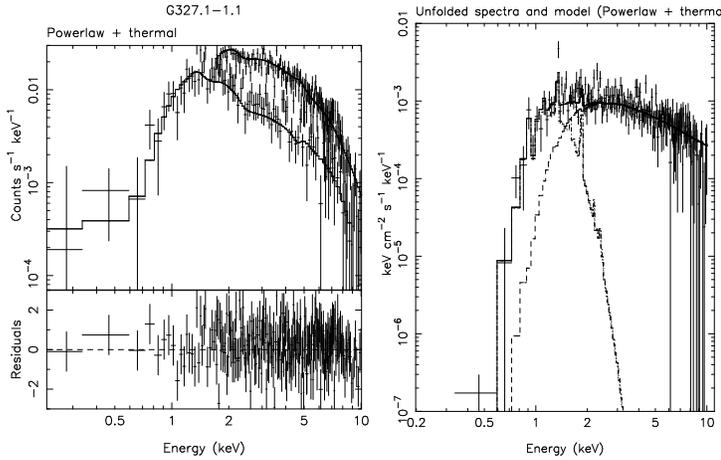

  \centerline{\hbox{
    \psfig{file=FIGURES/g327_pha.ps,width=6.0cm,angle=-90}
    \psfig{file=FIGURES/g327_euf.ps,width=6.0cm,angle=-90}
  }}
  \caption{Folded ({\rm left}) unfolded ({\rm right}) spectrum of the
  of \src, with best-fit power-law + thermal models.}
  \label{sp}
\end{figure}

\begin{figure}
  \centerline{\psfig{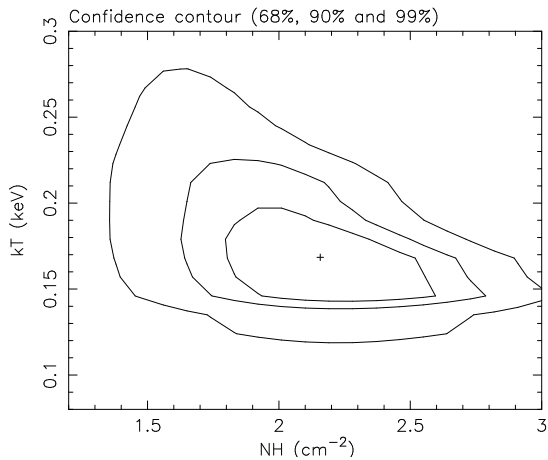}}
  \caption{Confidence (1-$\sg$, 2-$\sg$ and 3-$\sg$ level)
  contours in the $\nh$-$\Teff$ parameter plane
  derived for the total fit (Fit E in Table 1)
  to the LECS+MECS spectrum of \src.}
  \label{confcont}
\end{figure}

We have then performed a set of fits using a larger extraction region
(Fit D and E), that is $6\arcmin$ for MECS and $8\arcmin$ for LECS. The
aim of this selection is to include as much as possible the
contribution of the thermal component. The response has been generated
consistently, assuming a putative size for the shell, estimated using
the corrected LECS image in the 0.1--1 keV band (Fig.~\ref{shell}--top)
and applying the same procedure as that described for the plerionic
component: The measured, PSF and deconvolved size are respectively
$21.3\arcmin$, $5.1\arcmin$ and $19.7\arcmin$ FWHM.  The results of
these fits are also reported in Table~\ref{fitres}. Again the
two-component fit (Fit E) is statistically to be preferred to the pure
power-law one (Fit D): the $\chi^2$ decrease between the two is
significant at the 99.9\% confidence level.  The two-component fit of this spectrum
is shown in Fig.~\ref{sp}, while the contour levels for the
determination of the temperature and interstellar absorption are
reported in Fig.~\ref{confcont}.  The contribution of the thermal
component to the observed spectrum is low. We have estimated that the
thermal contribution to the total absorbed 0.5--10~keV flux is $\sim 10$\%,
and decreases to 0.4\% in the 2--10~keV band.

When compared with previous X-ray spectral analyses, we find a lower value for
the temperature and a higher value of $N_H$:
\citet{skr96} in fact derived $kT=0.8\pm0.3\U{keV}$ and
$\nh=1.3\pm0.4\E{22}\U{\cmq}$, while SWC gave
$kT=0.37^{+0.35}_{-0.20}\U{keV}$ and $\nh=1.8\pm0.3\E{22}\U{\cmq}$.
When the quoted uncertainties are taken into account, our results are still
marginally consistent with those of \citet{skr96} and SWC.
The discrepancy should be ascribed to differences in the
instrumental response of LECS (used here) and ROSAT PSPC (used both by SWC and
by \citet{skr96} for the soft band).

\begin{table*}
\caption{Summary of spectral fitting results.}
\label{fitres}
\medskip
\centering\begin{minipage}{13.3cm}
\begin{tabular}{lcccccc} \hline
Name & $\nh$ & $\gamma$ & 0.5--10 keV flux\footnote{Unabsorbed
non-thermal flux in the 0.5--10 keV band. For Fit A, C, D it refers to
the non-thermal component; while for Fit B and E we have reported the
flux of the thermal component only.} & kT & $\chi^2/dof$ \\

     &$10^{22}$ cm$^{-2}$ &  & $10^{-11}$ erg cm$^{-2}$ s$^{-1}$ & keV \\
\hline

Fit A - Ple 3\arcmin & $1.7^{<2.1}_{>1.4}$ & $2.11^{<2.25}_{>1.98}$ & 
$1.14^{<1.47}_{>0.92}$ & - & 139/160 \\
Fit B - Ple 3\arcmin & $2.4^{<3.2}_{>1.7}$ & $2.27^{<2.47}_{>2.08}$ & 
$63^{<7100}_{>0.9}$ & $0.16^{<0.33}_{>0.03}$ & 131/158 \\
Fit C - Ple 3\arcmin \ 2--10 keV & $2.3^{<3.1}_{>1.7}$ & $2.29^{<3.09}_{>1.68}$
 & $1.07^{<1.57}_{>0.80}$ & - & 118/141 \\
Fit D - G327 6\arcmin & $0.8^{<1.0}_{>0.6}$ & $1.94^{<2.07}_{>1.83}$ & 
$1.39^{<1.70}_{>1.18}$ & - & 235/232 \\
Fit E - G327 6\arcmin & $2.1^{<2.7}_{>1.4}$ & $2.25^{<2.44}_{>2.07}$ & 
$68^{<705}_{>3.0}$ & $0.18^{<0.22}_{>0.14}$ & 223/230 \\

\end{tabular}
\end{minipage}
\end{table*}

\subsection{The image of the thermal component}

\begin{figure}
  \centerline{\epsfig{file=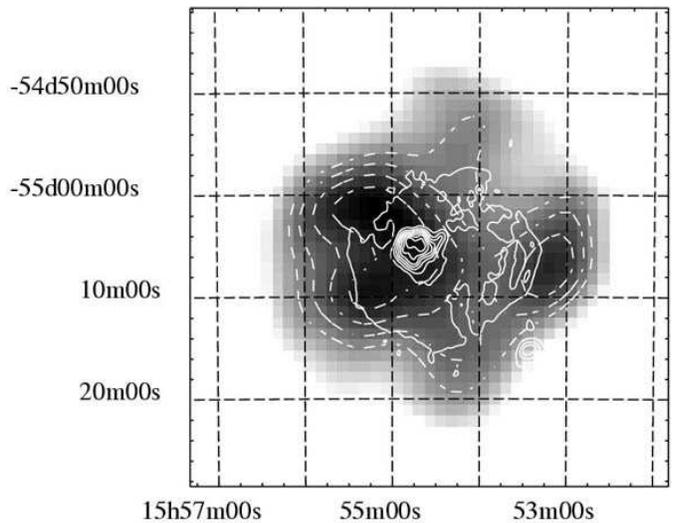,width=9.0cm}}
  \caption{0.1\dash1 keV image of the thermal component obtained by
  subtracting a scaled and smoothed MECS image from the total LECS
  image in the same band.  The 843 MHz total intensity radio contours
  of \protect\citet{wg96} are overlaid (solid), as well as surface
  brightness contours at 20\%, 40\%, 60\% and 80\% of the peak value.
  The thermal emission extends beyond the plerion in the NE and SE
  direction.}

  \label{shell}
\end{figure}

In order to further disentangle the image of the thermal component from
the total image, in a low-energy image we have subtracted the plerionic
component, by using the following approach.  First, we have extracted
the MECS image in the 2\dash10~keV energy band, in which the
contribution of the thermal component is far below the one of the
plerion.  Then we have convolved that image in order to simulate the
effect of the LECS PSF at $\sim1$~keV (we have found appropriate a
convolution with a gaussian of 5\arcmin\ FWHM).  Using the plerion
best-fit model (Fit B) reported in Table~\ref{fitres}, we have scaled
its image in order to reproduce the LECS image of the plerion only at
0.1\dash1~keV.  Finally, we have subtracted it from the original LECS
image of \src\ in the 0.1\dash1~keV band.  The result is shown in
Fig.~\ref{shell}.

The image is composed of a large diffuse component located in the
eastern part of the radio shell and a small X-ray blob located near
the western rim.  The center of the large diffuse component is located
at R.A.=$15\hh55\mm04.1\sss$, Dec.=$-55\degr05\arcmin08\arcsec$,
and it is displaced by $\sim 5.5\arcsec$ from
the radio SNR center and by $\sim 5.5\arcsec$ from the plerion center.
The observed size of this component is $\sim 21\arcmin$ FWHM.  An
interesting point is that the image of the thermal component does not
show a complete shell structure. It could be interpreted as partial
shell SNR, with a possible extension of the X-ray emission in the inner
region.  The absence of a limb can be inferred also from a PSPC image
of the whole X-ray SNR, as obtained by \citet{skr96}.


\subsection{Timing analysis}

We have searched for pulsation in the events collected in the region of
the plerion defined above. We have found no pulsation at the 99\%
confidence limit, with an upper limit at the same level of confidence
of 19.4\% in the 0.1--256 Hz frequency range.

\section{Discussion}

\subsection{The thermal component}

SWC have discussed a scenario in which the shell component of \src\
is expanding adiabatically in a homogeneous medium, and have thoroughly used
for it the well known \citet{sed59} analytic solution.
In this way SWC obtained a self-consistent scenario according to
which the remnant is $1.1\E4$~yr old and expands adiabatically in an ambient
medium with a density of $0.1\U{\cmc}$.

The two substantially new facts with respect to the starting point of their
analysis are that the X-ray structure cannot now be interpreted as a complete
shell, and that the effective temperature turned out
to be lower than previously determined.
Moreover, with respect to SWC we use a slightly higher estimate of
the average diameter of the SNR, namely 17\farcm5, as derived from the size of
the radio shell (\citealt{wg96}).

Let us first re-evaluate some parameters of \src\  assuming a \citet{sed59}
scenario.
In this way we derive how the difference from SWC in the starting point is going to
affect the result.
The basic relations for the effective temperature ($\Teff$), the age of the SNR
($\tSNR$), the emission measure parameter ($\EM$) defined as
$(\int n_en_i\,dV)/(4\pi d^2)$, the radius of the shock
($\Rsh$) and the total swept mass ($\Mtot$) are respectively:
\begin{eqnarray}
k\Teff&=&1.28k\,\Tsh=0.14\,\mH\Vsh^2,\\
\tSNR &=&0.4\,\Rsh/\Vsh,\\
\EM   &=&0.75\,\nz^2\Rsh^3/d^2,\\
\Rsh  &=&1.15\,(\ESN\tSNR^2/\rho_0)^{1/5},\\
\Mtot &=&4.19\,\rho_0\Rsh^3,
\end{eqnarray}
where $\ESN$ is the supernova energy, $\nz$ is the atomic number density in the
ambient medium (with standard abundances;
$\rho_0=1.26\,\mH \nz$), $d$ indicates the SNR distance,
$\Vsh$ and $\Tsh$ are respectively the shock velocity and the post-shock
temperature: the shock has been modelled assuming it to be a strong
non-radiative shock with temperature equilibration between ions and electrons.

Using the information on \src\ angular size, the above relations translate into
the following ones:
\begin{eqnarray}
\Vsh &=&814\,\TkeV^{1/2}\U{\kms},\\
\tSNR&=&12,200\,\TkeV^{-1/2}d_{10}\xi\U{yr},\\
\nz  &=&0.511\,\EM_{14}^{1/2}d_{10}^{-1/2}\xi^{-3/2}\U{\cmc},\\
\ESN &=&10.70\,\E{51}\TkeV\EM_{14}^{1/2}d_{10}^{5/2}\xi^{3/2}\U{erg},\\
\Mtot&=&1,100\,\EM_{14}^{1/2}d_{10}^{5/2}\xi^{3/2}\,\Msun,\\
\dE  &=&3.88\,E_{51}^{2/5}\EM_{14}^{-1/5}\TkeV^{-2/5}\xi^{-3/5}\U{kpc},
\end{eqnarray}
which directly link physical and observable quantities.
In these equations $\TkeV$ is the effective temperature, in keV;
$\EM_{14}$ is the emission measure parameter, in units of $10^{14}\U{cm^{-5}}$;
$d_{10}$ is the \src\ distance, in units of 10~kpc; and $E_{51}$ is the
supernova energy, when assumed, in units of $10^{51}\U{erg}$.
The quantity $\xi$ indicates the SNR angular size, in units of the \src\ one.
The physical quantities as derived from our spectral fit in Sedov approximation
are listed in the first column of Table~\ref{snrsedov}.
It is apparent that, while
the swept mass is much larger than any reasonable
stellar mass, the SNR shock is slow, so that the resulting supernova
energy is rather normal for any reasonable distance value.

\begin{table}
\caption{SNR parameters and associated $2\sg$ uncertainties 
derived by the fit to the Sedov and WL model.}
\label{snrsedov}
\medskip
\centering\begin{minipage}{9.0cm}
\begin{tabular}{lll}
\hline
        		& Sedov	model & WL model   \\
        		&             & ($C/\tau=3.25$)  \\
\hline
$\Vsh$ $(\!\U{\kms})$		& $345(304 - 430)$ &$590(470 - 790)$		\\
$\tSNR/d_{10}$ $(10^4\U{yr})$	& $2.9(2.6-3.3)$ & $1.7(1.2-2.1)$ \\
$\nz d_{10}^{1/2}$ $(\!\U{\cmc})$	& $0.37(0.10 - 1.35)$ &$0.11(0.02 - 0.41)$	\\
$E_{51} d_{10}^{-5/2}$	& $1.4(0.3 - 6.2)$ &$3.8(0.7 - 25.1)$	\\
$\Mx d_{10}^{-5/2}$ $(\Msun)$\footnote{For the Sedov model $\Mx=\Mtot$}
                    & $800(220-2910)$ & $1120(320-4000)$\\
$\dE$ $(\!\U{kpc})$	& $8.8(4.8 - 16.2)$ & $6.0(3.7-11.0)$\\
$\cooleff d_{10}^{-1/2}$ & $0.09(0.009 - 1.06)$ & $0.003(0.0001-0.14)$ \\
\end{tabular}
\end{minipage}
\end{table}

In other terms, assuming the standard value
of $10^{51}\U{erg}$ for the supernova energy, we derive a distance between 4.8 and 16.2 kpc, which is in agreement with published distance
estimates, ranging from 6.5~kpc (\citealt{skr96}) to 9~kpc (SWC),
as well with the distance estimated from the $\Sigma$-$D$ empirical relation
(\citealt{cb98}).
The average 1~GHz surface brightness of the \src\ shell component is
$2.4\E{-21}\U{W\,m^{-2}\, Hz^{-1} \,sr^{-1}}$ (5.6~Jy at 843~GHz, with a
spectral index $\sim-0.4$; SWC): this translates into a shell diameter of
$45\pm20\U{pc}$, and then into a distance estimate of $8.8\pm4.0$~kpc.

To summarize the results of our Sedov fit compared to those of SWC (using the
same assumed distance of 9~kpc), we find a SNR $\sim$2.4 times older and with
$\sim$4.7 times more energy, expanding in a medium $\sim$3.9 times denser.
The swept mass is 12.5 times larger than that derived by SWC.

\begin{figure}
\centerline{\psfig{file=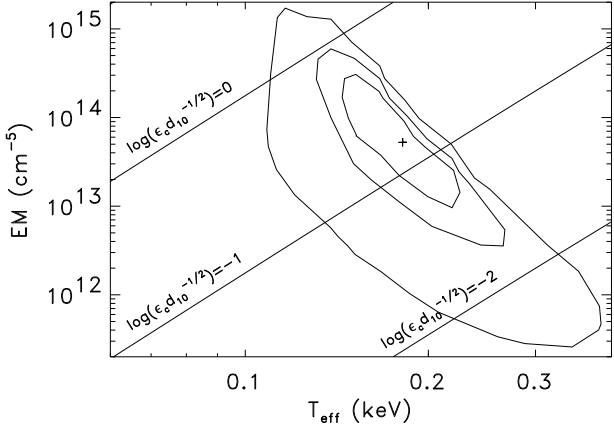,width=9.0cm}}
\caption{Efficiency of radiative cooling ($\cooleff$, scaled with
$d_{10}^{1/2}$), evaluated using Sedov solutions.
The solid straight lines show levels of the radiative
efficiency in the $\TkeV$-$\EM$ plane.
The best fit solution ($\Teff=0.18\U{keV}$; $\EM=0.52\E{14}\U{cm^{-5}}$) is
indicated by a cross.
The confidence contours are at 1-$\sg$, 2-$\sg$ and 3-$\sg$ level
(computed from Fit E in Table~\protect\ref{fitres}).}

\label{cool}
\end{figure}

The last row in Table~\ref{snrsedov} gives the efficiency in radiative
cooling, defined as the ratio between the SNR age and the present cooling time
right behind the shock, which can be expressed in terms of the observed
quantities as:
\begin{equation}
\cooleff=2.8\E{-3}(\TkeV^{-2.2}+0.33\,\TkeV^{-1})\sqrt{\EM_{14}d_{10}/\xi},
\end{equation}
where for the cooling coefficient we have used the formula
$\Lambda(T_6)=1.0\E{-22}T_6^{-0.7}+2.3\E{-24}T_6^{0.5}\U{erg\,cm^3\,s^{-1}}$,
with $T_6$ in units of $10^6\U{K}$ (\citealt{mcr87}).
A value for $\cooleff$ much smaller than unity is required in order to test the
adiabaticity condition, as required by the Sedov model.
However Fig.~\ref{cool} shows that,
since $\cooleff$ is strongly decreasing with $T$, fits
giving low values for $\Teff$ also imply relevant radiative losses.

The possible extension of the thermal X-ray component to the inner
shell regions, together with the non-negligible radiative cooling,
leads us to consider scenarios alternative to the Sedov one for the
evolution of \src.

Let us first examine the WL model, describing (by a class of self-similar
solutions) a wide range of conditions under which a SNR may expand in a cloudy
interstellar medium.
In this scenario, a delayed evaporation of the clouds causes
the central density to increase with respect to the pure Sedov case.
This class of solutions depends on two parameters: the mass fraction in clouds
($C$) and the cloud evaporation time scaled with the SNR age ($\tau$).
The solutions converge to the Sedov case when the mass deposition is small,
either because little mass is contained in the clouds ($C\ll1$) or because
the clouds
have anyway released little mass during the SNR lifetime ($C\ll\tau$).
Here we assume that the mass in clouds is large ($C\gg1$), but the evaporation
time is long ($\tau\gg1$): in this case the solutions form a one-parameter
class, depending on just $C/\tau$ (these solutions are approximately valid also
for $C,\tau\ga1$).


We have qualitatively compared the radial profile of the thermal
component (using the center coordinates reported in Sect. 3.2) with the
WL predicted profiles for various value of the $C/\tau$ parameter.  We
found a reasonable matching, considering the uncertainties involved,
for $C/\tau$ in the range between 2.5 and 4.0.  These uncertainties
originate not only from the faintness of the source, but also from
deviations from the assumed spherical symmetry as well as from a non
perfect subtraction of the plerionic component: all of these effects,
with the present data, are even difficult to quantify.  Therefore,
under the assumption that WL models are appropriate to model the
\src\ thermal component, we can estimate $C/\tau=3.25\pm0.75$. By using
the results tabulated by WL we derive which correction factors must be
applied to the Sedov estimates.

For instance
\begin{equation}
\bRsh=\Rsh/\RshS=	(0.81\pm0.04)
\end{equation}
shows how much the shock radius shrinks with respect to the Sedov case (for
fixed $\ESN$, $\nz$ and $\tSNR$): with reference to the quantity $K$, defined in
Eq.~7 of WL, the relation $\bRsh^5=K/K\rs{S}$ holds.
Moreover the equations
\begin{eqnarray}
\bEM  &=&\EM/\EMS=	(10\pm8);\\
\bTeff&=&\Teff/\TeffS=	(0.33\pm0.12);\\
\bMx  &=&\Mx/\MtotS=	(4.5\pm1.6)
\end{eqnarray}
refer to comparisons with the Sedov solution keeping $\nz$ and $\Rsh$
constant. The above scaling factors are evaluated by interpolating data from
Table~6 of WL. In the above expressions we have used the index $\rs{S}$ to
indicate the values in the Sedov solution.

Therefore the corrections required to derive, for the WL model, the relations
equivalent to Eqs.~6--11 for the associated Sedov case are:
\begin{eqnarray}
\Vsh &=&\bTeff^{-1/2}\VshS= (1.7\pm0.3)\,\VshS,\\
\tSNR&=&\bTeff^{1/2}\tSNRS= (0.6\pm0.1)\,\tSNRS,\\
\nz  &=&\bEM^{-1/2}\nzS= (0.3\pm0.1)\,\nzS,\\
\ESN &=&\bRsh^{-5}\bEM^{-1/2}\bTeff ^{-1}\ESNS= (2.7\pm0.7)\,\ESNS,\\
\Mx  &=&\bMx\bEM^{-1/2}\MtotS= (1.40\pm0.04)\,\MtotS,\\
\dE  &=&\bRsh^2\bEM^{1/5}\bTeff^{2/5}\dES= (0.68\pm0.07)\,\dES:
\end{eqnarray}
It should be noted that here by ``associated'' WL and Sedov solutions we
indicate those showing the same observed properties (i.e.\ size, $\Teff$ and
$\EM$).

The values of the parameters derived with the WL model are also given in
the second column of Table~\ref{snrsedov}.
Compared to the pure Sedov case, now the ambient
(intercloud) density results to be lower, while the shock velocity is higher
(which also makes the SNR younger).  The distance estimate based on the
$10^{51}\U{erg}$ energy requirement now gives $\sim6.0\U{kpc}$, still
compatible with the lowest distance estimates for this source
(\citealt{skr96}).  The cooling efficiency $\cooleff$ (in the
intercloud medium) can be evaluated from Eq.~12, by substituting $\TkeV$ with
$\TkeV/\bTeff$ and $\EM_{14}$ with $\EM_{14}/\bEM$:
it results much smaller ($\sim3\E{-2}$) than in the Sedov case.

On the basis of our analysis, a WL model fits the average radial profile of the
Thermal
remnant better than the Sedov model.
Moreover the WL model gives reasonable values of the SNR parameters, even
though: 1. it leads to a rather small SNR age, with problems for the offset of
the associated neutron star (see below); 2. for a standard SNR energy, it leads
to a distance somehow smaller than what estimated by other methods.

A further scenario which may account for thermal X-ray from inner regions
is that of a SNR in radiative phase (with in addition possible
effects, like thermal conduction and flux saturation). A common property of
non-radiative spherical shocks expanding in a homogeneous medium is the
presence in the inner region of a hot and thin medium, which therefore behaves
as a very low efficiency emitter, and forms the typical hollow sphere
emission pattern of shell-type SNRs. In a cloud evaporation model (as WL, see
above), this region is filled with material originally stored in small clouds.
But since inside the shell the pressure equilibrium is roughly maintained, it
would be sufficient to cool down this gas (even without adding further material)
in order to make it denser.

Unfortunately no self-similar solution is known for radiative shell-type SNRs,
and therefore a quantitative description of this class of objects requires to
develop numerical models, which are beyond the scope of this work. Models of
this kind have been already developed, for instance for W44, by various
authors (\citealt{hhs97,csm99,scm99}). Here let us simply compare the basic
parameters of \src\ and W44 to infer their relative evolutive conditions,
under the assumption that both have reached the radiative phase.

Table~\ref{srcw44} lists, for both remnants, some basic parameters, either
observed or derived: in the latter case a Sedov model has been used for the
derivation. Thus if these SNRs are no longer in Sedov phase, the listed values
are
inaccurate (even though, simply for dimensional arguments, not too far from the
correct values).

For instance the absolute ages, as reported in Table~\ref{srcw44}, may be
understimated since in radiative phase the SNRs must have decelerated with
respect to the Sedov phase. We know this is the case for W44, where the Sedov
estimate is about half of what derived from more detailed NEI models, as
well as half of the pulsar spin-down age (\citealt{hhs97}).

If also \src\ is in radiative phase, its true age may be substantially larger
than 29,000~yr (as from Table~\ref{srcw44}). At any rate, if we exclude the WL scenario, this SNR must be much
older than originally estimated by SWC (11,000~yr). An implication would be that the
lower limit on the velocity of the compact object ($>600$ km s$^{-1}$, as from SWC,
in order to account for the displacement of the compact source from the SNR
center)
can be lowered by a factor 3 or even more: therefore in this case there is
no need for an anomalous pulsar velocity.

Although some values listed in Table~\ref{srcw44} may be inaccurate in an
absolute sense, we are confident that they may still be used for a comparative
analysis
of the two SNRs. For instance \src\ looks older than W44; it also expands
with a lower velocity, and in a slightly thinner medium. Moreover, as shown in
the last line of the table, the cooling efficiency in \src\ is much
higher than in
W44, and this fact strengthens the idea that, if a radiative scenario results
to be appropriate 
to the case of W44, it should be even more reasonable for \src.

\begin{table}
\caption{Comparison between \src\ and W44 (from \citealt{hhs97}) parameters 
(estimated using a Sedov model); for W44 we used Eqs.~6 through 12 with 
$\xi=1.76$.}
\label{srcw44}
\medskip
\begin{tabular}{lll}
\hline
		&\src\				&W44			\\
\hline
$\Rsh$		&$25.5\,d_{10}\U{pc}$		&$11.2\,d_{2.5}\U{pc}$	\\
$kT$	&$0.18\U{keV}$			&$0.88\U{keV}$		\\
$\nsav V$	&$7.0\E{59}d_{10}^2\U{\cmc}$	&$1.3\E{58}d_{2.5}^2\U{\cmc}$\\
$\Lrad$		&$6.2\E{32}d_{10}^2\U{\ergs}$	&$1.7\E{33}d_{2.5}^2\U{\ergs}$\\
\hline
age		&$2.9\E4d_{10}\U{yr}$		&$5.7\E3d_{2.5}\U{yr}$	\\
$\nz$		&$0.37d_{10}^{-1/2}\U{\cmc}$	&$0.18d_{2.5}^{-1/2}\U{\cmc}$\\
$\cooleff$	&$0.09d_{10}^{1/2}$		&$7.5\E{-4}d_3^{1/2}$		\\
\noalign{\smallskip}
\hline
\end{tabular}
\end{table}

Although expanding in a higher ambient density, 
\src\ radio shell is about 3 times fainter than W44: also this difference
could be a consequence of its age.
The fact that high resolution maps of the radio continuum from W44 present a
strongly filamented pattern has been taken (\citealt{csm99}) as evidence for
the remnant being in radiative phase.
If this is the case, higher resolution radio maps of \src, able to test the
presence of filamentation, could be valuable to shed light on the evolutive
phase of this remnant.

\subsection{The plerionic component}

Like W44, \src\ shows both a shell and a plerionic component in
radio, while a thermal filled-centre and a non-thermal component in X rays: the
main difference is that in \src\ the plerionic component is more prominent than
in W44.
Another similarity between W44 and \src\ is the
displacement of the plerionic component from the shell centroid, as well as
its bow shock shape.

An analysis of the integrated spectrum of the plerionic component can be
carried on in a very similar way as already done by SWC. Our estimate of the
0.5--10~keV luminosity of the plerion ($\sim1.6\E{35}d_{10}^2\U{\ergs}$) is
about 0.5 times that estimated by SWC; moreover the age we find is at least a
factor 2.4 larger, and all this affects the derivation of pulsar and nebular
parameters.

By applying an empirical relation (\citealt{sw88}),
SWC used the X-ray nebular luminosity ($L_X$) to infer the pulsar spin down
power ($\dot E$), which for a magnetic dipole braking is proportional to $\dot
P/P^3$. The age estimate ($\tSNR \sim P/2\dot P$) is the other relation which allows
one separating the pulsar timing parameters $P$ and $\dot P$, and to estimate
also the pulsar surface field: SWC report $P=62\U{ms}$ and $B_0=2.3\E{12}\U{G}$.
Since by this procedure $P\propto \tSNR^{-1/2}\dot E^{-1/2}$,
while $B_0\propto \tSNR^{-1}\dot E^{-1/2}$, our revised values lead
to a somehow shorter period ($<50\U{ms}$) and to a lower surface field
($<1.2\E{12}\U{G}$).
The presence of a faster, and lower-B pulsar compared to that in W44 is in fact
required by the presence of a reasonably bright, although aged, plerion.

SWC also used a spectral break in the nebular spectrum (at about
$3.5\E{4}\U{GHz}$) to infer a nebular field $B\sim0.7\E{-4}\U{G}$. Since the
field derived in this way is $\propto \tSNR^{-2/3}$,
we infer instead a nebular field $\la0.4\E{-4}\U{G}$. For a radius of
$\sim1.5\U{arcmin}$ for the (radio) plerion, the total magnetic energy is
$\la6\E{47}\U{erg}$.
The plerion is then slightly underpressured ($P_B<0.2\E{-10}\U{dyn\,cm^{-2}}$),
compared to the thermal remnant, if in Sedov or WL regime; while a better
pressure balance may be attained if the shell is in radiative expansion.
Anyway the uncertainties involved in these estimates are too large to take
this as a strong argument in favour of a radiative shell.

\section{Summary and conclusions}

We have used a \sax\ X-ray observation of \src\ to study the origin of
its X-ray emission, with particular emphasis on its thermal component,
which has not been studied in detail in the literature, because of
its superposition with the bright non-thermal plerion. We fitted the
plerion and the total SNR spectrum with a combination of thermal
and non-thermal models, finding an X-ray temperature $kT\sim 0.18$
keV for the hot plasma and a power-law photon index $\sim 2.2$ for
the plerion. Using the flux derived for the non-thermal plerion, we
have subtracted the non-thermal component from the total 0.1--1 keV SNR
image, thus obtaining for the first time an image of the thermal hot
plasma.

The emission from the plasma is not
confined to a bright shell, and shows possible emission for inner regions.
We have then examined for this SNR also scenarios different from the Sedov one.
The model of
SNR expansion in a medium with evaporative clouds developed by
\citet{wl91} may account for the observations, 
although it may imply an age
and distance for this object which are somehow smaller than expected.
We have
shown that a radiative expansion with strong central cooling may also
describe the data. We have derived and compared the remnant
characteristic parameters for the various scenarios, either Sedov, or
\citet{wl91}, or
radiative model, showing that the last model implies a very long age
($\ga 3\times 10^4$ yr): in this case the observed offset of the associated
pulsar may be accounted for without requiring a too high spatial
velocity for it.

A comparison with W44, a remnant for which there is also evidence of
radiative expansion, indicates that \src\ may be substantially more
evolved than W44. Future high sensitivity X-ray and radio observations
are strongly encouraged, because they may dissipate any doubt about
the evolutionary stage of this object and in particular of its shell.
Finally, we have also revised the estimate of the pulsar and nebular
magnetic field.

\begin{acknowledgements}

F. Bocchino acknowledges partial support from MIUR and ASI.
This work has been partly supported by MIUR under Grant Cofin2002--02--10.

\end{acknowledgements}

\bibliographystyle{aa}
\bibliography{references}


\end{document}